# Stabilization of Rydberg Dissipative Time Crystals Using a Scanning Fabry–Pérot Interferometer Transfer Lock

D. Arumugam, B. Feyissa

*Jet Propulsion Laboratory, California Institute of Technology*

(Electronic mail: darmindra.d.arumugam@jpl.nasa.gov.)

(Dated: 30 September 2025)

Stabilization of laser frequencies is critical for sensitive Rydberg measurements, including in applications such as dissipative time-crystal (DTC) dynamics, yet conventional approaches often require complex or costly hardware. We demonstrate a compact, low-cost stabilization method using a scanning Fabry–Pérot interferometer (SFPI) to transfer lock a 960nm coupler laser to an 852nm probe. The lock suppresses coupler multi-MHz free-running drift and improves the Allan deviation by up to an order of magnitude, reaching <75kHz at $\tau\sim$66s. Applied to DTC oscillations using a Rb 2-photon D2 transition, the second harmonic generated 480nm (from 960nm lock) reduces DTC frequency drift from >20kHz to a few kHz and lowers instability by more than an order of magnitude with a minimum Allan deviation of 0.2kHz at $\tau<$10s. These results establish SFPI-based transfer locking as a practical and accurate approach for scalable multi-laser Rydberg experiments requiring long-term stability in a compact and low-cost system.

Precise laser frequency stabilization is essential for atomic and Rydberg experiments, since drift and noise directly degrade spectroscopic resolution, sensitivity, and long-term stability [1]. Several methods are widely used. Direct atomic locking relies on Doppler-free absorption or modulation spectroscopy [2,3]. Cavity-based schemes such as Pound–Drever–Hall (PDH) achieve sub-kHz stability but require vibration-isolated optical cavities in ultra-high-vacuum [4]. Recently, digital and FPGA-based controllers have enabled flexible feedback with cavities/atomic transitions over wide dynamic range [5,6]. Each approach balances complexity, cost, and performance, but all aim to ensure lasers remain as stable as the quantum systems under study.

Rydberg atoms provide a powerful platform for quantum science and sensing, enabling strong interaction studies, nonlinear optics, and atom-based electrometry [7–9]. Applications now range from RF receivers to deployable field sensors, driving rapid growth in the field [10–12]. In typical two-photon schemes, the probe is locked to an atomic hyperfine line, while the coupler drives a Rydberg transition at wavelengths without direct atomic references. Electromagnetically induced transparency (EIT) can be used for coupler stabilization [13–15], but EIT locks require continuous atomic monitoring, scale poorly to multi-laser operation, and consume valuable coupler power, reducing what is available for experiments. For emerging portable Rydberg sensors there is thus a clear need for compact, low-cost locking approaches. Conventional methods such as PDH cavity locks or modulation spectroscopy remain powerful but impose alignment, cost, and environmental burdens, motivating the development of simpler alternatives.

In this article, we demonstrate a scanning Fabry–Pérot interferometer (SFPI) transfer lock for stabilizing the coupler laser in a two-photon Rydberg system. This capability enables robust observation of dissipative time-crystal (DTC) oscillations, a many-body dynamical phase of interest for quantum science and low-frequency field sensing [16–18]. Rydberg-based DTCs are promising for portable electrometry in the sub-MHz regime [19,20]. Using the SFPI lock, adapted from recent scalable transfer-cavity approaches [21], we suppress multi-MHz coupler drift and reduce Allan deviation by more than an order of magnitude, reaching <75kHz at $\tau\sim$66s. We show that DTC oscillation frequencies remain stable within a few kHz with a minimum Allan deviation of 0.2kHz at $\tau<$10s. These results establish SFPI stabilization as a compact, low-cost method for Rydberg experiments.

I. ATOMIC SYSTEM AND EXPERIMENT

The experiments are performed in a room-temperature vapor cell of enhanced Rb-87 (length 56 mm, diameter 25 mm, see Fig. 1). A pair of Helmholtz coils provides a uniform bias B-field along the propagation axis of the probe and coupler beams, allowing control of Zeeman shifts. The atomic ladder system is formed by the $5S_{1/2}\rightarrow 5P_{3/2}\rightarrow 63D_{3/2}$ excitation pathway. A weak probe laser at 780nm addresses the $5S_{1/2},F=2\rightarrow 5P_{3/2},F'=3, |g\rangle\rightarrow|e\rangle$ cyclic transition, while the coupler laser is generated by frequency-doubling a 960nm diode laser to 480nm, resonant with the $5P_{3/2}\rightarrow 63D_{3/2}$ Rydberg $|e\rangle\rightarrow|r\rangle$ transition and detuned by $\Delta c/2\pi$= +22MHz. The probe laser is frequency-stabilized by Doppler-free saturation spectroscopy in a separate Rb reference cell, providing a probe linewidth <82kHz. For the coupler, the 960nm fundamental is stabilized to an 852nm Cs (Cesium) reference laser using the SFPI (Finesse ≥ 1500, 824-1071nm, 1.5 GHz FSR). The SFPI is built as a rigid cavity assembly containing two parallel, high-reflectivity mirrors with a clear 9.5mm aperture. The mirror separation is fixed at 50mm, and the cavity is mounted on a piezoelectric transducer (PZT) for fine scanning of the optical path length. Direct use of the 780nm probe with the SFPI was not possible, as the cavity mirrors are coated for >824nm. The Cs 852nm was frequency-stabilized by standard Doppler-free saturation spectroscopy. The stabilized 960nm light is then frequency-doubled using a resonant second-harmonic generation (SHG), a nonlinear crystal embedded in a bow-tie high finesse cavity to generate the 480nm radiation delivered to the vapor cell. Typical optical powers are 180μW for the probe at 780nm and 320mW at 480nm for the coupler. Both beams have diameters of ~1mm (1/e²) and aligned in a counter-propagating configuration through the cell. Quarter-wave plates are used to adjust both probe and coupler ellipticity to maximize the signal-to-noise ratio of the oscillatory response (see Fig. 1c) [18-20]. The transmitted probe is collected on a fast photodiode and recorded on both a digital oscilloscope and spectrum analyzer. Additional RF band-pass filter (1–240 kHz, 3 dB cutoffs) is applied electronically to suppress technical noise outside the frequency band of the dissipative oscillations.

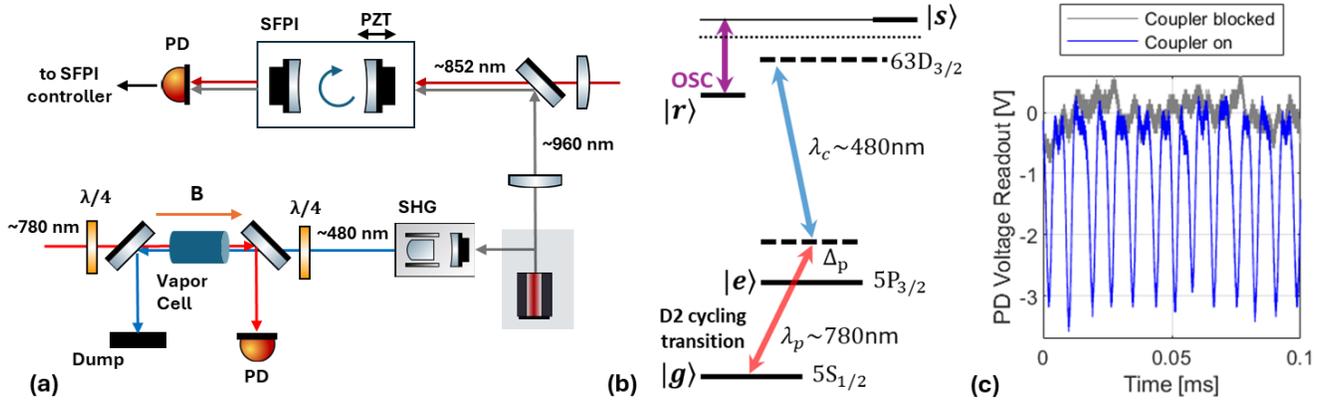

**Figure 1.** SFPI transfer lock and stabilized Rydberg dissipative time-crystal oscillations. (a) Setup: the reference 852-nm probe is locked to the Cs hyperfine line, and the coupler is stabilized relative to the reference using a scanning Fabry–Pérot interferometer (SFPI) before driving Rb $63D_{3/2}$. (b) Level scheme: the probe excites $5S_{1/2} \rightarrow 5P_{3/2}$, and the SFPI-locked coupler drives $5P_{1/2} \rightarrow$ Rydberg $63D_{3/2}$, enabling limit-cycle oscillations. (c) Photodiode readout: with the coupler blocked (gray) only background noise is present, while with the coupler on and locked (blue) robust oscillations are observed.

Under these conditions, EIT is observed in probe transmission as the coupler is tuned across the resonance. The DTC dynamics manifest as sustained oscillations between the driven Rydberg state $|r\rangle$ and nearby sublevels $|s\rangle$, with the magnetic bias B-field providing careful control of the Zeeman splitting and the probe–coupler polarizations enforcing the selection rules that stabilize these oscillations [16-20]. We use a B-field magnitude of ~4G. At higher coupler intensities, the system undergoes a Hopf bifurcation, achieving self-sustained oscillations in the transmitted probe — the hallmark of a DTC phase (example in [16-20]). As shown in Fig. 1c, oscillations are only sustained when the coupler is present; with the coupler blocked, only background detector noise remains. The DTC oscillation frequencies are typically in the 100–200 kHz range with the present setup, with amplitude and linewidth strongly dependent on coupler detuning and optical power. Optical powers for both probe and coupler are stabilized by individual noise eaters (laser amplitude stabilization) using liquid-crystal retarders driven by servos. Long-duration frequency stability of these DTC oscillations requires frequency locking of the coupler laser.

## II. LOCKING SCHEME

Stabilization of the coupler laser at 960nm is achieved using a digital feedback loop based on SFPI, which references the coupler to a stable 852nm probe laser (see Fig. 2a). The SFPI provides a free spectral range of 1.5 GHz and finesse >1500, allowing simultaneous observation of both the coupler and reference peaks within each cavity scan with a <1MHz resolution. The transmitted light is detected with a photodiode, amplified, and digitized. A synchronized trigger from the FPI controller ensures that each scan is recorded with fixed timing.

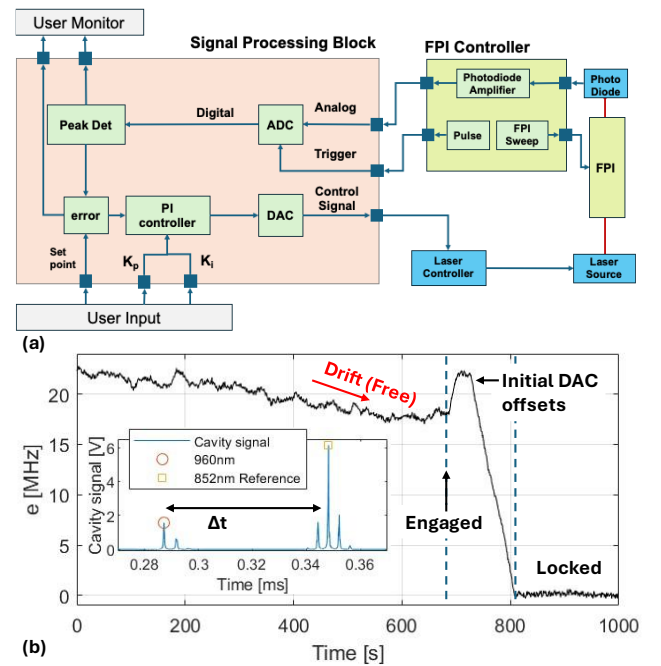

**Figure 2.** Digital SFPI lock architecture and drift suppression. (a) Block diagram of the feedback loop for stabilizing the coupler laser. The Fabry–Pérot interferometer (FPI) signal is detected on a photodiode, digitized, and synchronized to the cavity sweep. FFT peak detection yields the frequency offset between the 960-nm coupler and 852-nm probe reference. The error is processed by a PI controller, converted by a DAC, and applied to the laser, with user control of lock point and gain. (b) Frequency drift of the coupler. In the free-running case (left), the offset varies steadily. When the SFPI lock is engaged at ~700s, initial DAC transients settle into a stable lock with no further drift. Inset: cavity transmission showing the time separation Δt between 960-nm and reference 852-nm peaks used to generate the error signal.

During each scan, the cavity transmission peaks corresponding to the 852nm and 960nm lasers are identified in software. The time separation Δt between the peaks is proportional to their frequency offset and is used as the error signal. This error is processed by a proportional–integral (PI) controller with adjustable gains ($K_p$, $K_i$), converted to an analog correction signal via a DAC (digital to analog convertor), and applied to the coupler laser controller to correct drift. The digital feedback loop operated at a repeat rate of 25.4ms per scan, corresponding to a correction bandwidth of ~40Hz.

Figure 2b shows the lock suppressing the drift. In the free-running case, the coupler frequency drifts steadily by many MHz over several hundred seconds. At ~700 s, the lock is engaged, introducing a short DAC transient before the system converges to a stable locked state with no further drift. The inset illustrates a typical cavity scan, highlighting the measured Δt between the 960nm coupler and 852nm reference peak that defines the error relative to a set-point.

## III. RESULTS

The effect of the SFPI-based stabilization on the coupler laser is summarized in Fig. 3. In the free-running case (red), the coupler exhibits large frequency excursions of several megahertz within two hundred seconds. Such drift is typical for diode-based sources and can severely limit the stability of multi-photon Rydberg experiments. When the SFPI lock is engaged, the drift is strongly suppressed (see Fig. 2b): the locked trace (blue in Fig. 3a for frequency drift) remains confined within a narrow band around zero detuning with no long-term wander. This improvement is further quantified by the Allan deviation in Fig. 3b. While the free-running case shows steadily increasing instability with averaging time, reaching >0.6MHz beyond τ ≥280s, the SFPI-locked coupler achieves more than an order-of-magnitude reduction, with Allan deviation below 75 kHz at τ ≥66s.

At a coupler power of P=320mW focused to a 1mm diameter beam (w = 0.5mm), the peak intensity is $I_0$ = $2P/(\pi w^2) \approx 8.1 \times 10^5$ W/m², giving an electric field amplitude $E_0 = \sqrt{2I_0/(c\epsilon_0)} \approx 2.5 \times 10^4$ V/m. Using the $m_J$-averaged dipole matrix element for the $5P_{3/2} \to 63D_{3/2}$ transition ($|d| \approx 2.2 \times 10^{-32}$ C·m, corresponding to ~$2.6 \times 10^{-3} ea_0$), the coupler Rabi frequency is $\Omega_c = |d|E_0/\hbar \approx 5.2 \times 10^6$ rad/s, or $\Omega_c/2\pi \approx 0.83$MHz. The expected two-photon EIT linewidth in this ladder system is then $\Delta\nu_{EIT} \approx \gamma_{rg} + \Omega_c^2/\gamma_e$, with $\gamma_e/2\pi \approx 6$ MHz (5P natural width) and $\gamma_{rg}/2\pi \sim 0.3$–0.5 MHz (estimated effective ground–Rydberg decoherence rate), yielding $\Delta\nu_{EIT} \approx$ 0.4–0.6MHz. The relevant stability metric is the Allan deviation on experimental timescales, not absolute long-term drift. Our SFPI-based lock achieves an Allan deviation of <75kHz at 960nm (i.e. <150kHz at the 480nm SHG) for averaging times τ ≥ 66s, corresponding to ≤30% of the two-photon linewidth.

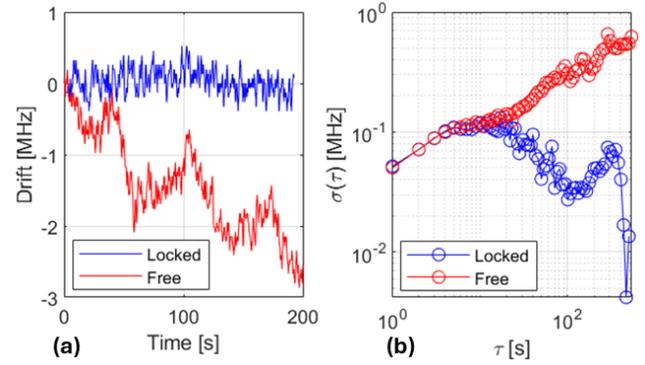

**Figure 3.** Drift suppression of the 960-nm coupler laser with the SFPI lock. (a) Time-domain frequency drift relative to the 852-nm reference. The free-running coupler (red) shows multi-MHz excursions over 200 s, while the SFPI-locked case (blue) remains confined near zero. (b) Allan deviation. The locked coupler (blue) achieves a reduction in instability by more than an order of magnitude relative to the free-running case (red) for averaging times of τ>20–60s.

The impact of the coupler stabilization on DTC oscillations is illustrated in Fig. 4. In the free-running case (Fig. 4a,b), the oscillation peak appears between 140-160kHz with pronounced spectral spreading (Fig. 4a, blue) when averaged over time (200s), reflecting the frequency spread of the DTC oscillation. This effect is also visible in the spectrogram, where the DTC oscillation frequency drifts continuously. By contrast, with the SFPI lock engaged (Fig. 4c,d), the oscillation peak shifts to ~125 kHz and maintains well-defined narrow spectrum (Fig. 4d). The suppression of frequency drift eliminates the broadening observed in the free-running spectrum (blue lines in Fig. 4a vs Fig. 4c). These results confirm that locking the coupler laser stabilizes the DTC oscillation frequency, improving long-term measurements without degradation of DTC oscillation signal fidelity.

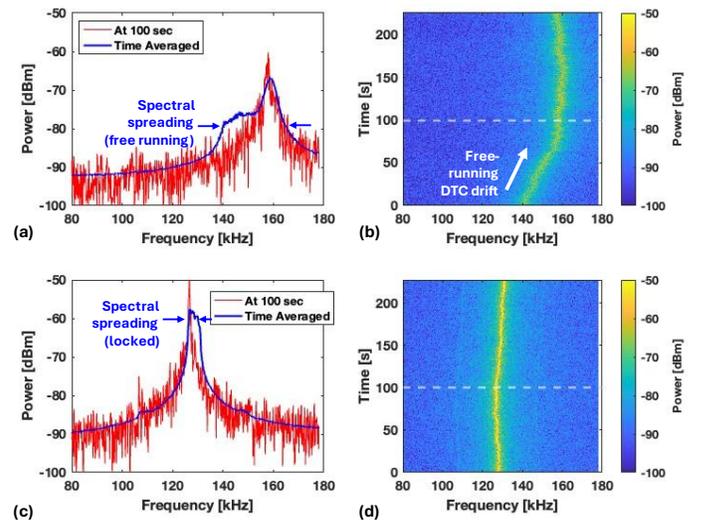

**Figure 4.** Frequency stability of dissipative time-crystal oscillations with and without SFPI locking. (a,b) Free-running coupler laser: the oscillation peak near 140-160 kHz shows frequency wander, visible in both the spectra (a) and the time–frequency spectrogram (b). (c,d) SFPI-locked coupler: the oscillation peak near 125 kHz remains narrow and centered, with reduced spectral wandering in both the spectra (c) and spectrogram (d). Locking suppresses long-term drift, yielding stable oscillation frequency and improved reproducibility.

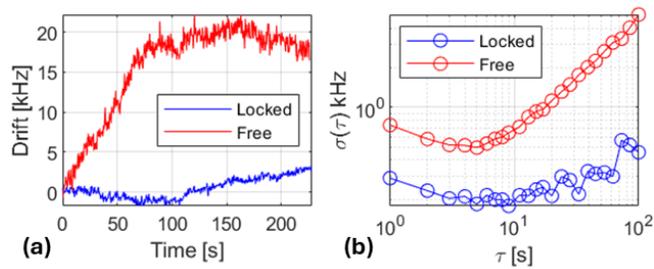

**Figure 5.** Oscillation frequency stability with and without SFPI locking. (a) Time-domain drift of the oscillation frequency. In the free-running case (red) the frequency drifts by >20 kHz over 200 s, whereas the locked case (blue) remains confined within a few kHz. (b) Allan deviation of the oscillation frequency. Locking (blue) reduces instability by more than an order of magnitude across averaging times of τ>20s compared to the free case (red), demonstrating effective stabilization of the dissipative time-crystal oscillation frequency.

Figure 5 quantifies the stability of DTC oscillations. In the free-running case (red, Fig. 5a), the oscillation frequency drifts by >20kHz over 200s, whereas with the SFPI lock (blue) the drift remains within only a few kilohertz. The Allan deviation (Fig. 5b) shows more than an order-of-magnitude improvement for $\tau \gtrsim 20$s, with the locked oscillations reaching Allan deviations at the sub-kilohertz level (minimum Allan deviation of 0.2kHz at $\tau<$10s). These results confirm that Rydberg coupler stabilization via a compact (5-7cm scale) and cost-effective (total SFPI cost <$4.2k) SFPI scheme directly translates into robust long-term stability of the DTC oscillation frequency.

ACKNOWLEDGMENTS

The research was carried out at the Jet Propulsion Laboratory, California Institute of Technology, under a contract with the National Aeronautics and Space Administration (80NM0018D0004), through the Instrument Incubator Program's (IIP) Instrument Concept Development (Task Order 80NM0022F0020).

AUTHOR CONTRIBUTIONS

D.A. proposed the project. D.A. configured the quantum systems to include lasers, SFPI, and stabilization/locking systems. D.A. and B.F developed the digital system for SFPI locking and laser error reduction. D.A. developed and optimized excitation and field sources for DTC structure observation. D.A. and B.F. designed the software scripts for data collection and processed the data for the figures. Both authors supported data collection efforts. Both authors contributed to discussions of the results and the manuscript.

DATA AVAILABILITY STATEMENT

The frequency drift of the coupler from free-running transitioning to locked in Fig. 2b is available as source data. All other data is available upon reasonable request to the corresponding author, Darmindra Arumugam via email: darmindra.d.arumugam@jpl.nasa.gov.

[1]H. Metcalf and P. van der Straten, *Laser Cooling and Trapping* (Springer, 1999).

[2]E. Arimondo, "Coherent population trapping in laser spectroscopy," *Prog. Opt.* **35**, 257–354 (1996).

[3]J. L. Hall, "Nobel Lecture: Defining and measuring optical frequencies," *Rev. Mod. Phys.* **78**, 1279 (2006).

[4]R. W. P. Drever et al., "Laser phase and frequency stabilization using an optical resonator," *Appl. Phys. B* **31**, 97–105 (1983).

[5]J. Appel et al., "A versatile digital laser frequency lock for spectroscopy and cooling of atoms," *Rev. Sci. Instrum.* **87**, 063104 (2016).

[6]R. Hobson, W. Bowden, and P. Gill, "FPGA-based digital laser locking," *Rev. Sci. Instrum.* **85**, 114705 (2014).

[7]T. F. Gallagher, Rydberg Atoms (Cambridge Univ. Press, 1994).

[8]J. A. Sedlacek et al., "Microwave electrometry with Rydberg atoms in a vapor cell," Nat. Phys. 8, 819–824 (2012).

[9]H. Fan, S. Kumar, J. Sedlacek, H. Kübler, S. Karimkashi, and J. P. Shaffer, "Atom based RF electric field sensing," J. Phys. B 48, 202001 (2015).

[10]D. A. Anderson, S. A. Miller, and G. Raithel, "Rydberg atoms for radio-frequency communications and sensing: atomic receivers," IEEE Trans. Antennas Propag. 69, 2455–2467 (2021).

[11]J. Kuebler et al., "Rydberg atom sensors for RF field measurement," Appl. Phys. Lett. 116, 194001 (2020).

[12]P. Sedlacek, H. Kübler, R. Löw, and T. Pfau, "Rydberg-atom electrometry for space and field applications," Appl. Phys. B 127, 2 (2021).

[13]M. Fleischhauer, A. Imamoglu, and J. P. Marangos, "Electromagnetically induced transparency: optics in coherent media," Rev. Mod. Phys. 77, 633–673 (2005).

[14]A. K. Mohapatra, T. R. Jackson, and C. S. Adams, "Coherent optical detection of highly excited Rydberg states using EIT," Phys. Rev. Lett. 98, 113003 (2007).

[15]M. Tanasittikosol et al., "Microwave electrometry with Rydberg atoms using EIT," J. Phys. B 44, 184020 (2011).

[16]Xiaoling Wu, Zhuqing Wang, Fan Yang, Ruochen Gao, Chao Liang, Meng Khoon Tey, Xiangliang Li, Thomas Pohl, and Li You, "Dissipative time crystal in a strongly interacting Rydberg gas," Nat. Phys. (2024). https://www.nature.com/articles/s41567-024-02542-9

[17]K. Wadenpfuhl and C. S. Adams, "Emergence of synchronization in a driven-dissipative hot Rydberg vapor," Phys. Rev. Lett. 131, 143002 (2023). https://doi.org/10.1103/PhysRevLett.131.143002

[18]Dongsheng Ding et al., "Ergodicity breaking from Rydberg clusters in a driven-dissipative many-body system," Sci. Adv. 10, eadl5893 (2024). https://doi.org/10.1126/sciadv.adl5893

[19]D. Arumugam, "Electric-field sensing with driven-dissipative time crystals in room-temperature Rydberg vapor," Scientific Reports 15, 13446 (2025).

[20]D. Arumugam, "Stark-modulated Rydberg dissipative time crystals at room-temperature applied to sub-kHz electric-field sensing," arXiv:2503.08972 (2025).

[21]E. Pultinevicius et al., "A scalable scanning transfer cavity laser stabilization scheme based on the Red Pitaya STEMlab platform," Rev. Sci. Instrum. 94, 103004 (2023). https://doi.org/10.1063/5.0169021